\documentclass[twocolumn]{aastex631}

\shorttitle{Multi-wavelength emission from MADs around isolated black holess}
\shortauthors{Kimura, Kashiyama, Hotokezaka}
\graphicspath{{./}{figures/}}

\begin{document}

\title{Multi-wavelength emission from magnetically arrested disks around isolated black holes}

\correspondingauthor{Shigeo S. Kimura}
\email{shigeo@astr.tohoku.ac.jp}

\author[0000-0003-2579-7266]{Shigeo S. Kimura}
\altaffiliation{JSPS Fellow}
\affiliation{Frontier Research Institute for Interdisciplinary Sciences, Tohoku University, Sendai 980-8578, Japan}
\affiliation{Astronomical Institute, Graduate School of Science, Tohoku University, Sendai 980-8578, Japan}

\author[0000-0003-4299-8799]{Kazumi Kashiyama}
\affiliation{Department of Physics, Graduate School of Science, University of Tokyo, Bunkyo-ku, Tokyo 113-0033, Japan}
\affiliation{Research Center for the Early Universe, Graduate School of Science, University of Tokyo, Bunkyo-ku, Tokyo 113-0033, Japan}
\affiliation{Kavli Institute for the Physics and Mathematics of the Universe (Kavli IPMU,WPI), The University of Tokyo, Chiba 277-8582, Japan}

\author[0000-0002-2502-3730]{Kenta Hotokezaka}
\affil{Research Center for the Early Universe, Graduate School of Science, University of Tokyo, Bunkyo-ku, Tokyo 113-0033, Japan}



\begin{abstract}
We discuss the prospects for identifying nearest isolated black holes (IBHs) in our Galaxy. 
IBHs accreting gas from the interstellar medium (ISM) likely form magnetically arrested disks (MADs). 
We show that thermal electrons in the MADs emit optical signals through the thermal synchrotron process while non-thermal electrons accelerated via magnetic reconnections emit a flat-spectrum synchrotron radiation in the X-ray to MeV gamma-ray ranges.
The Gaia catalog will include at most a thousand of IBHs within $\lesssim 1$ kpc that are distributed on and around the cooling sequence of white dwarfs (WDs) in the Hertzsprung-Russell diagram.
These IBH candidates should also be detected by eROSITA, with which they can be distinguished from isolated WDs and neutron stars.
Followup observations with hard X-ray and MeV gamma-ray satellites will be useful to unambiguously identify IBHs. 
\end{abstract}

\keywords{Stellar mass black holes(1611), Compact radiation sources(289), Non-thermal radiation sources(1119), Accretion(14), Plasma astrophysics(1261)}


\section{Introduction} \label{sec:introduction}

The existence of stellar-mass black holes (BHs) is confirmed by dynamical motion in X-ray binaries \citep{2016ApJS..222...15T,2016A&A...587A..61C} and gravitational-wave detection \citep{LIGOScientific:2020ibl}. Stellar-mass BHs are believed to form as an end product of stars of initial masses higher than $\sim$ 25 $M_\odot$ \citep{WHW02a}. Considering the star-formation rate and age of the Universe, there should be roughly $10^8$ BHs in our Galaxy \citep[e.g.,][]{2020ApJ...905..121A}, suggesting that the nearest BH should be located $\lesssim50$ pc from the Earth. However, only a few tens of BHs have been discovered, most of which are in X-ray binaries and located $\gtrsim1$ kpc away from the Earth \citep{2016A&A...587A..61C,2016ApJS..222...15T}. Vast majority of stellar-mass BHs in our Galaxy wandering in the interstellar medium (ISM) have yet to be identified.

Wandering BHs, or isolated BHs (IBHs), accrete gas of the ISM via Bondi-Hoyle-Littleton accretion \citep{Edg04a}, and accretion flows should be formed around the IBHs. The accretion flows emit multi-wavelength signals, and detection prospects of these signals have been discussed for a long time with various methods and assumptions  \citep{1975A&A....44...59M,1985MNRAS.217...77M,1998ApJ...495L..85F,2002MNRAS.334..553A,2003ApJ...596..437C,2012MNRAS.427..589B,IMT17a,2018MNRAS.477..791T,2019MNRAS.488.2099T}.

In this Letter, we discuss prospects to identify nearest IBHs, newly considering two effects. One is the multi-wavelength emission model of magnetically arrested disks (MADs; \citealt{NIA03a,MTB12a}). MADs are expected to be formed when the mass accretion rate onto the BH is significantly lower than the Eddington rate  \citep{2011ApJ...737...94C,2021ApJ...915...31K}, and typical IBHs accrete ISM gas with a highly sub-Eddington rate \citep{IMT17a}.
Thermal electrons are heated up to a relativistic temperature by dissipation of magnetic energy \citep{2018MNRAS.478.5209C,2021MNRAS.506..741M}, and they emit optical signals through thermal synchrotron radiation. MADs also accelerate non-thermal electrons via magnetic reconnections \citep{2012SSRv..173..521H,2020PhPl...27h0501G}, which produce X-rays and MeV gamma-rays through synchrotron radiation \citep{BOP16a,2020MNRAS.494.5923P,2021arXiv210708056S,2021arXiv210915115R}.

The other is to consider the prospects for detection by Gaia \citep{2016A&A...595A...1G} and eROSITA \citep{2021A&A...647A...1P}. These satellites will provide complete catalogs of Galactic objects more than ever before, and they  likely contain accreting IBHs. In order to distinguish IBHs from other objects, we need to understand multi-wavelength spectra of accreting IBHs and develop a strategy for identifying them. We will describe the multi-wavelength emission model of MADs around IBHs (IBH-MADs), and show how IBH-MADs can be distinguishable from other astronomical objects. We use convention of $Q_X=Q/10^X$ in cgs unit except the BH mass for which we use $M_X=M/10^X M_\odot$.

\section{IBH-MAD model} \label{sec:MADs}

\begin{table}[t]
\begin{center}
\caption{Physical quantities in 5 ISM phases. $n_{\rm ISM}$, $C_{s,\rm ISM}$, $H_{\rm ISM}$, $\xi_0$ are the number density, effective sound velocity, scale height, and volume filling factor of the ISM phases. We mainly discuss Cold HI, Warm HI, and Warm HII. }\label{tab:ism}
 \begin{tabular}{|c|cccc|}
\hline
ISM phase & $n_{\rm ISM}$ & $C_{s,\rm ISM}$ & $H_{\rm ISM}$ & $\xi_0$ \\
 &  [cm$^{-3}$] &  [km s$^{-1}$] &  [kpc] & \\
\hline
Molecular clouds & $10^2$ & 10 & 0.075 & 0.001 \\
{\bf Cold HI} & $10$ & 10 & 0.15 & 0.04 \\
{\bf Warm HI} & 0.3 & 10 & 0.50 & 0.35 \\
{\bf Warm HII} & 0.15 & 10 & 1.0 & 0.2\\
Hot HII & 0.002 & 150 & 3.0 & 0.43\\
\hline
 \end{tabular}
\end{center}
\end{table}

Accretion rates onto IBHs strongly depend on the physical properties of the ISM and IBH. We consider five-phase ISM given by \cite{2000eaa..bookE2636B}, which is also used in the  literature \citep[e.g.,][]{2002MNRAS.334..553A,IMT17a,2018MNRAS.477..791T}. The physical parameters characterizing each ISM phase is tabulated in Table \ref{tab:ism}. We find that Gaia can detect IBHs in hot HII medium only when they are extremely close ($d\lesssim10$ pc) and/or massive ($M\gtrsim40~M_\odot$). Also, Gaia may be unable to measure the intrinsic color of IBHs in molecular clouds due to strong dust extinction, and thus it is difficult to identify IBHs in molecular clouds (but see e.g., \citealt{2018MNRAS.475.1251M}). Hence, we hereafter focus on the other three phases.

We estimate the physical properties of IBH-MADs.
Since the accretion rate is much lower than the Eddington rate, $\dot{M}_{\rm Edd}= L_{\rm Edd}/c^2\simeq 1.4\times10^{18}M_1\rm~g~s^{-1}$, the radiatively inefficient accretion flow (RIAF; \citealt{Ich77a,ny94,YN14a}) is formed.
According to recent general relativistic magnetohydrodynamic (GRMHD) simulations, RIAFs can produce outflows and create large-scale poloidal magnetic fields even starting from purely toroidal magnetic field \citep{2020MNRAS.494.3656L}. These poloidal fields are efficiently carried to the IBH, which likely results in formation of a MAD around the IBH \citep{2011ApJ...737...94C,IMT17a,2021ApJ...915...31K}\footnote{Some GRMHD simulations do not achieve the MAD state even for their long integration timescales, depending on the initial magnetic field configurations \citep{2012MNRAS.426.3241N,2020ApJ...891...63W}. This may indicate that the condition for MAD formation depends on the magnetic field configurations of the ambient medium.}.
Introducing a reduction parameter of the mass accretion rate, $\lambda_w\le1$, due to outflows and convection \citep{BB99a,2000ApJ...539..809Q,2015ApJ...804..101Y,2018MNRAS.476.1412I},
the accretion rate onto an IBH can be estimated as
\begin{eqnarray}
 \dot{M}_\bullet&\approx&\lambda_w\frac{4\pi G^2M^2\mu_{\rm ISM}m_pn_{\rm ISM}}{\left(C_s^2+v_k^2\right)^{3/2}}\\
 &\simeq&7.3\times10^{10}\lambda_{w,0}M_1^2n_{\rm ISM,-1}\left(\frac{\sqrt{C_s^2+v_k^2}}{40\rm~km~s^{-1}}\right)^{-3}\rm~g~s^{-1},\nonumber
\end{eqnarray}
where $G$ is the gravitational constant, $M$ and $v_k$ are the mass and the proper-motion velocity of the IBH, respectively, $m_p$ is the proton mass, and $\mu_{\rm ISM}\simeq1.26$, $n_{\rm ISM}$, and $C_s$ are the mean atomic weight, number density, and sound speed of the ISM gas (see Table \ref{tab:ism}), respectively. We use $\lambda_w=1$ as a reference value for simplicity, but we will discuss the cases with a low value of $\lambda_w$ in Section \ref{sec:discussion}. We assume $v_k\simeq40\rm~km~s^{-1}$ as a reference value as in \citet{IMT17a}

The radial velocity, proton temperature, gas number density, and magnetic field of MADs can be estimated to be \citep{2019MNRAS.485..163K,Kimura:2020thg,2021ApJ...915...31K}
\begin{eqnarray}
 V_R&\approx& \frac12\alpha V_K\simeq1.5\times10^9\mathcal{R}_1^{-1/2}\alpha_{-0.5}\rm~cm~s^{-1} ,\\
 k_BT_p&\approx&\frac{GMm_p}{4R}\simeq23\mathcal{R}_1^{-1}\rm~MeV\\
  N_p&\approx&\frac{\dot{M}_\bullet}{4\pi RHV_R\mu_{\rm ISM}m_p} \\
&\simeq&2.3\times10^{10}\dot{M}_{\bullet,11}M_1^{-2}\mathcal{R}_1^{-3/2}\alpha_{-0.5}^{-1}\rm~cm^{-3},\nonumber\\
 B&=& \sqrt{\frac{8\pi N_pk_BT_p}{\beta}}\\
 &\simeq&1.5\times10^4\dot{M}_{\bullet,11}^{1/2}M_1^{-1}\mathcal{R}_1^{-5/4}\alpha_{-0.5}^{-1/2}\beta_{-1}^{-1/2}\rm~G,\nonumber
\end{eqnarray}
where $\mathcal{R}=R/R_G$ is the size of the emission region normalized by the gravitational radius, $R_G=GM/c^2$, $\alpha$ is the viscous parameter \citep{ss73}, $H\approx R/2$ is the scale height, and $\beta$ is the plasma beta. 

Inside MADs, electrons are heated up to a relativistic temperature by magnetic energy dissipation, such as magnetic reconnections \citep{2017ApJ...850...29R,2018ApJ...868L..18H} and the turbulence cascades \citep{how10,2019PNAS..116..771K}.  We parameterize the total heating rate and electron heating rate as 
\begin{eqnarray}
Q_{\rm thrml}&=&\epsilon_{\rm dis}(1-\epsilon_{\rm NT})\dot{M}_\bullet c^2,\\
Q_{e,\rm thrml}&=&f_e Q_{\rm thrml}  \label{eq:Qe}\\
&\simeq&2.7\times10^{30}\left(\frac{f_e\epsilon_{\rm dis}(1-\epsilon_{\rm NT})}{0.3\cdot0.15\cdot0.67}\right)\dot{M}_{\bullet,11} \rm~erg~s^{-1},\nonumber
\end{eqnarray}
where $\epsilon_{\rm dis}$ is the ratio of dissipation to accretion energies, $\epsilon_{\rm NT}$ is the ratio of non-thermal particle production to dissipation energy, and $f_e$ the electron heating fraction. Considering the trans-relativistic magnetic reconnection, we use the electron heating prescription given by \cite{2017ApJ...850...29R,2018MNRAS.478.5209C}\footnote{Previous works on emissions from MADs \citep{2021ApJ...915...31K,2020ApJ...905..178K} use the prescription by \cite{2018ApJ...868L..18H}, which assumes non-relativistic magnetic reconnections. Since magnetic reconnections in MADs can be trans-relativistic, we examine \cite{2018MNRAS.478.5209C} in this study.}:
\begin{equation}
f_e\approx\frac12\exp\left(-\frac{1-4\beta\sigma_B}{0.8+\sqrt{\sigma_B}}\right),
\end{equation}
where $\sigma_B=B^2/(4\pi N_pm_pc^2)\simeq0.5\mathcal{R}_1^{-1}\beta_{-1}^{-1}$ is the magnetization parameter. We assume that the proton temperature is sub-relativistic, which is reasonable for the bulk of the accretion flows. We obtain $f_e\sim0.3$ with our reference parameter set.

\section{Photon spectra from IBH-MADs}\label{sec:spectrum}

We calculate the photon spectrum from IBH-MADs using the method in \cite{2021ApJ...915...31K} (see also \citealt{kmt15,2019PhRvD.100h3014K,2020ApJ...905..178K}), where we include both thermal and non-thermal components of electrons and treat them as separate components. Thermal electrons emit broadband photons by thermal synchrotron, bremsstrahlung, and Comptonization processes. Non-thermal electrons emit broadband photons by synchrotron emission, and we can ignore other emission processes in the MADs. We also calculate emissions induced by non-thermal protons, but we find that their contribution is negligible.

  \begin{figure*}
   \begin{center}
    \includegraphics[width=\linewidth]{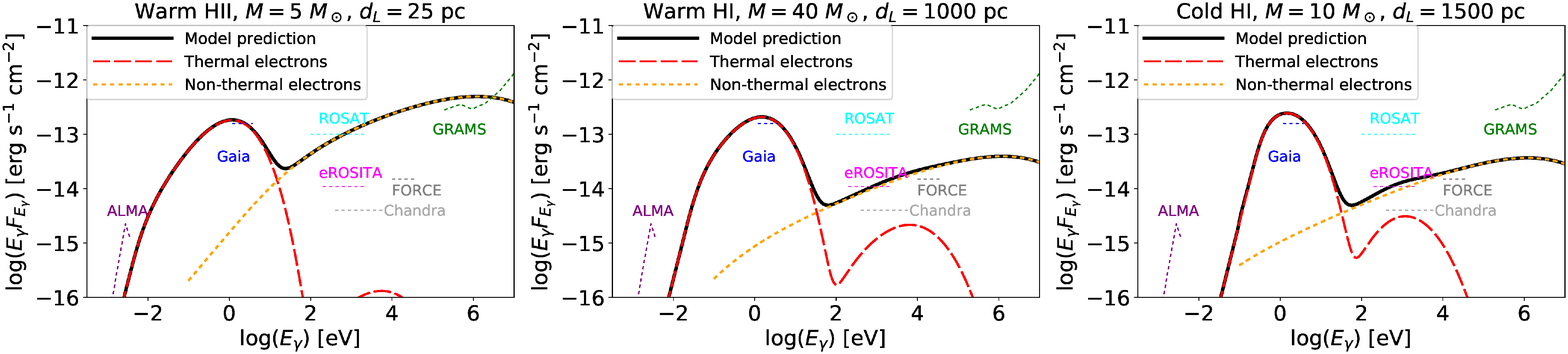}
    \caption{Broadband spectra from IBH-MADs. The thick-solid, thick-dashed, and thick-dotted lines are total, photon spectra by thermal electrons, and photon spectra by non-thermal electrons. The thin-dashed lines are sensitivity curves for ALMA (purple: 30 min; \href{https://almascience.nao.ac.jp/proposing/sensitivity-calculator}{ALMA Sensitivity Calculator}), Gaia (blue; 20 mag; \citealt{2018A&A...616A...1G}), eROSITA (magenta; 4-year survey; \citealt{2021A&A...647A...1P}), Chandra (gray; 10 ks; \href{https://cxc.harvard.edu/cdo/about_chandra/}{CXO website}), FORCE (darkgray; 100 ks; \citealt{2018SPIE10699E..2DN}), and GRAMS (green; 3 year; \citealt{2020APh...114..107A}). ISM phase, black hole mass, and distances are shown in each panel. Other parameters are $\mathcal{R}=10$, $\alpha=0.3$, $\beta=0.1$, $\epsilon_{\rm dis}=0.15$, $\epsilon_{\rm NT}=0.33$, $\eta_{\rm acc}=5$, and $s_{\rm inj}=1.3$. }
    \label{fig:spe}
   \end{center}
  \end{figure*}

The thermal electrons emit optical photons by thermal synchrotron radiation.
For cases with low $\dot{M}_\bullet$, the cooling processes are so inefficient that the radiative cooling cannot balance the heating before falling to the IBH. Then, the electron temperature is determined by $k_BT_{e,\rm adi}\approx f_ek_BT_p\simeq7.0(f_e/0.3)\mathcal{R}_1^{-1}$ MeV. For high $\dot{M}_\bullet$, the electron temperature is determined by the balance between the heating and cooling, i.e., $Q_{e,\rm thrml}=L_{\rm thrm}(T_{e,\rm rad})$, where $L_{\rm thrm}(T_{e,\rm rad})$ is the radiative cooling rate. The electron temperature in IBH-MADs are given by $T_e={\rm min}(T_{e,\rm adi},~T_{e,\rm rad})$.

Because of their lower accretion rate compared to quiescent X-ray binaries by 2-3 orders of magnitude, IBH-MADs are optically thin for synchrotron-self absorption (SSA) at the synchrotron peak frequency in the most parameter space. This feature is different from any other RIAF systems, such as quiescent X-ray binaries \citep{1996ApJ...457..821N,2021ApJ...915...31K}, radio galaxies \citep{2020ApJ...905..178K}, low-luminosity AGNs \citep{nse14,kmt15,2019PhRvD.100h3014K,Kimura:2020thg}, and Sgr A* \citep{NYM95a,mmk97,yqn03}\footnote{The Eddington ratio for Sgr A* is estimated to be lower than that for IBH-MADs. Nevertheless, the RIAF around Sgr A* is expected to be optically thick for SSA at the peak frequency because of its lower synchrotron peak frequency and larger emission region.}.
Since the optically thin thermal synchrotron emission has a gradual spectral cutoff \citep{1996ApJ...465..327M}, the peak frequency of the synchrotron spectrum is $\simeq25$ times higher\footnote{We can derive the factor 25 by taking derivative of Equation (36) in \cite{1996ApJ...465..327M}.} than the canonical synchrotron frequency, $\nu_{\rm syn}=3eB\theta_e^2/(4\pi m_ec)$, where $\theta_e=k_BT_e/(m_ec^2)$. Then, the peak frequency of the thermal synchrotron emission is estimated to be
\begin{equation}
 \nu_{\rm syn,pk}\approx\frac{75eB\theta_e^2}{4\pi m_ec}\simeq2.0\times10^{14}B_4\mathcal{R}_1^{-2}\left(\frac{f_e}{0.3}\right)^2\rm~Hz,
\end{equation}
where we use $T_e=T_{e,\rm adi}$.
The luminosity of the thermal synchrotron emission is roughly estimated to be
\begin{eqnarray}
& &\nu_{\rm syn,pk} L_{\nu_{\rm syn,pk}}\approx\frac43(3\theta_e)^2\frac{\sigma_TcB^2}{8\pi}(\pi R^3N_p) \label{eq:Lsyn}\\
&\simeq&9.3\times10^{28}\dot{M}_{\bullet,11}^2M_1^{-1}\mathcal{R}_1^{-3}\alpha_{-0.5}^{-2}\beta_{-1}^{-1}\left(\frac{f_e}{0.3}\right)^2\rm~erg~s^{-1}.\nonumber
\end{eqnarray}
Comparing Equations (\ref{eq:Qe}) and (\ref{eq:Lsyn}), the critical mass accretion rate above which the cooling is efficient can be estimated to be 
\begin{eqnarray}
\dot{M}_{\rm cl}\simeq2.9\times10^{12}M_1\mathcal{R}_1^3\alpha_{-0.5}^2\beta_{-1}\left(\frac{f_e}{0.3}\right)^{-1}\\
\times\left(\frac{\epsilon_{\rm dis}(1-\epsilon_{\rm NT})}{0.15\cdot0.67}\right)\rm~g~s^{-1}. \nonumber
\end{eqnarray}
With our reference parameters, typical IBH-MADs in the warm media are in the adiabatic regime, while those in the cold medium are in the cooling regime.

Magnetic reconnections accelerate non-thermal electrons which emit X-rays and soft gamma-rays by synchrotron radiation. We consider non-thermal particle injection, cooling, and escape processes, and solve the steady-state transport equation to obtain the number spectrum, $N_{E_e}$ (see \citealt{2020ApJ...905..178K,2021ApJ...915...31K} for details). The injection spectrum is assumed to be a power law with an exponential cutoff, i.e., $\dot{N}_{E_e,\rm ing}\propto E_e^{-s_{\rm inj}}\exp(-E_e/E_{e,\rm cut})$, where $E_{e,\rm cut}$ is the cutoff energy and $s_{\rm inj}$ is the injection spectral index. Although earlier 2D particle-in-cell (PIC) simulations result in a cutoff energy of $E_{\rm cut}\sim4\sigma_B$ \citep{2016ApJ...816L...8W}, long-term calculations revealed that the cutoff energy is increasing with time \citep{2018MNRAS.481.5687P,2021arXiv210500009Z}. Since the dynamical timescale of the accretion flow is much longer than the timescales of kinetic plasma phenomena, we determine $E_{\rm cut}$ by the balance between the acceleration and cooling processes. 
The injection rate is normalized by $\int\dot{N}_{E_e,\rm inj}E_edE_e=f_e\epsilon_{\rm NT}\epsilon_{\rm dis}\dot{M}_\bullet c^2$. We consider only the synchrotron cooling as the other processes are negligible.  We consider both advective (infall to the IBH) and diffusive escapes.  The acceleration time is phenomenologically set to be $t_{\rm acc}=\eta_{\rm acc}E_ec/(eBV_A^2)$, where $V_A=B/\sqrt{4\pi m_pN_p}$ is the Alfven velocity and $\eta_{\rm acc}$ is the acceleration efficiency parameter.

In the range of our interest, the synchrotron cooling limits the maximum energy, and the synchrotron cutoff energy is estimated to be 
\begin{equation}
E_{\gamma,\rm cut}\approx\frac{3e^2h_p\beta_A^2}{m_ec\sigma_T\eta_{\rm acc}}\simeq15\left(\frac{\beta_A}{0.7}\right)^2\left(\frac{\eta_{\rm acc}}{5}\right)^{-1}\rm~MeV.
\end{equation}
The peak luminosity for the non-thermal synchrotron process is roughly estimated to be $E_\gamma L_{E_\gamma}\approx f_e\epsilon_{\rm NT}\epsilon_{\rm dis}\dot{M}c^2$.
The cooling break energy is given by equating infall time to the cooling time: $E_{\gamma,\rm cl}\approx h_peB\gamma_{e,\rm cl}^2/(2\pi m_ec)\approx1.2\times10^2B_4\gamma_{e,\rm cl,3}^2$ eV, where $\gamma_{e,\rm cl}\approx{\rm max}(1,6\pi m_ecV_R/(\sigma_TB^2R))\simeq7.7\times10^2V_{R,9}B_4^{-2}R_7^{-1}$ is the cooling break Lorentz factor. The X-ray band is typically above the cooling break energy, and thus the photon index in the X-ray band is $\Gamma_X=(s_{\rm inj}+2)/2\simeq1.65$ with $s_{\rm inj}=1.3$  (see Section 5 for discussion on the value of $s_{\rm inj}$). Then, the X-ray luminosity is estimated to be
\begin{equation}
 E_XL_X\simeq1.3\times10^{29} \dot{M}_{\bullet,11}f_{X,-1}\left(\frac{f_e\epsilon_{\rm NT}\epsilon_{\rm dis}}{0.3\cdot0.33\cdot0.15}\right)\rm~erg~s^{-1},\label{eq:Lx}
\end{equation}
where $f_X=(E_X/E_{\gamma,\rm cut})^{2-\Gamma_X}\sim0.1$ is the correction factor. IBH-MADs in the adiabatic regime roughly exhibit $L_X/L_{\rm opt}\sim1$ with our reference parameters, as seen by Equations (\ref{eq:Lsyn}) and (\ref{eq:Lx}).
In the cooling regime of $\dot{M}_\bullet>\dot{M}_{\rm cl}$, both thermal and non-thermal electrons emit all the energies via synchrotron emission. Then, we can write $L_X/L_{\rm opt}\approx f_X\epsilon_{\rm NT}/(1-\epsilon_{\rm NT})\sim0.05$ with our reference parameters.

Figure \ref{fig:spe} shows the broadband photon spectra from IBH-MADs, whose parameters are shown in each panel and the caption. The parameters in our MAD model are calibrated using the gamma-ray data of radio galaxies \citep{2020ApJ...905..178K} and the multi-wavelength data of quiescent X-ray binaries \citep{2021ApJ...915...31K}. 
The thermal synchrotron emission produces optical signals that is detectable by Gaia. 
The synchrotron emission by non-thermal electrons produce power-law photons from X-ray to MeV gamma-ray ranges. The IBHs detectable by Gaia should be detected by eROSITA \citep{2021A&A...647A...1P}.
SSA is effective in radio and sub-mm bands, and thus, it is challenging to detect IBH-MADs by radio telescopes, such as ALMA (see Section \ref{sec:discussion} for radio signals from jets associated with IBH-MADs).
For a low accretion rate, the advection cooling is effective for thermal electrons, while the radiative cooling is efficient for non-thermal electrons. Then, emission by non-thermal electrons can be more luminous than that by thermal electrons, despite we choose $\epsilon_{\rm NT}<1-\epsilon_{\rm NT}$, as seen in the left panel of Figure \ref{fig:spe}.

\section{Strategy to identify IBHs}\label{sec:identify}

  \begin{figure*}
   \begin{center}
    \includegraphics[width=\linewidth]{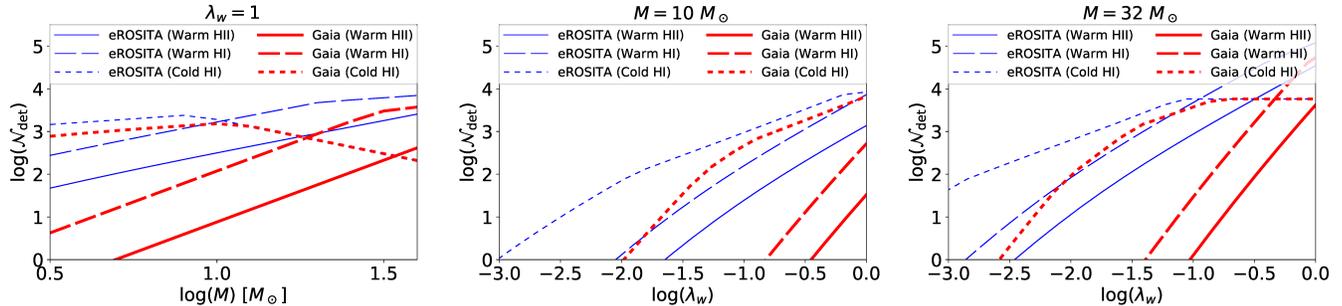}
    \caption{Expected numbers of IBHs detected by Gaia (thick-red) and eROSITA (thin-blue) as a function of $M$ (left) and $\lambda_w$ (middle and right) in various ISM phases. The solid, dashed, and dotted lines are for warm HII, warm HI, and cold HI, respectively.  }
    \label{fig:dndm}
   \end{center}
  \end{figure*}

  \begin{figure}
   \begin{center}
    \includegraphics[width=\linewidth]{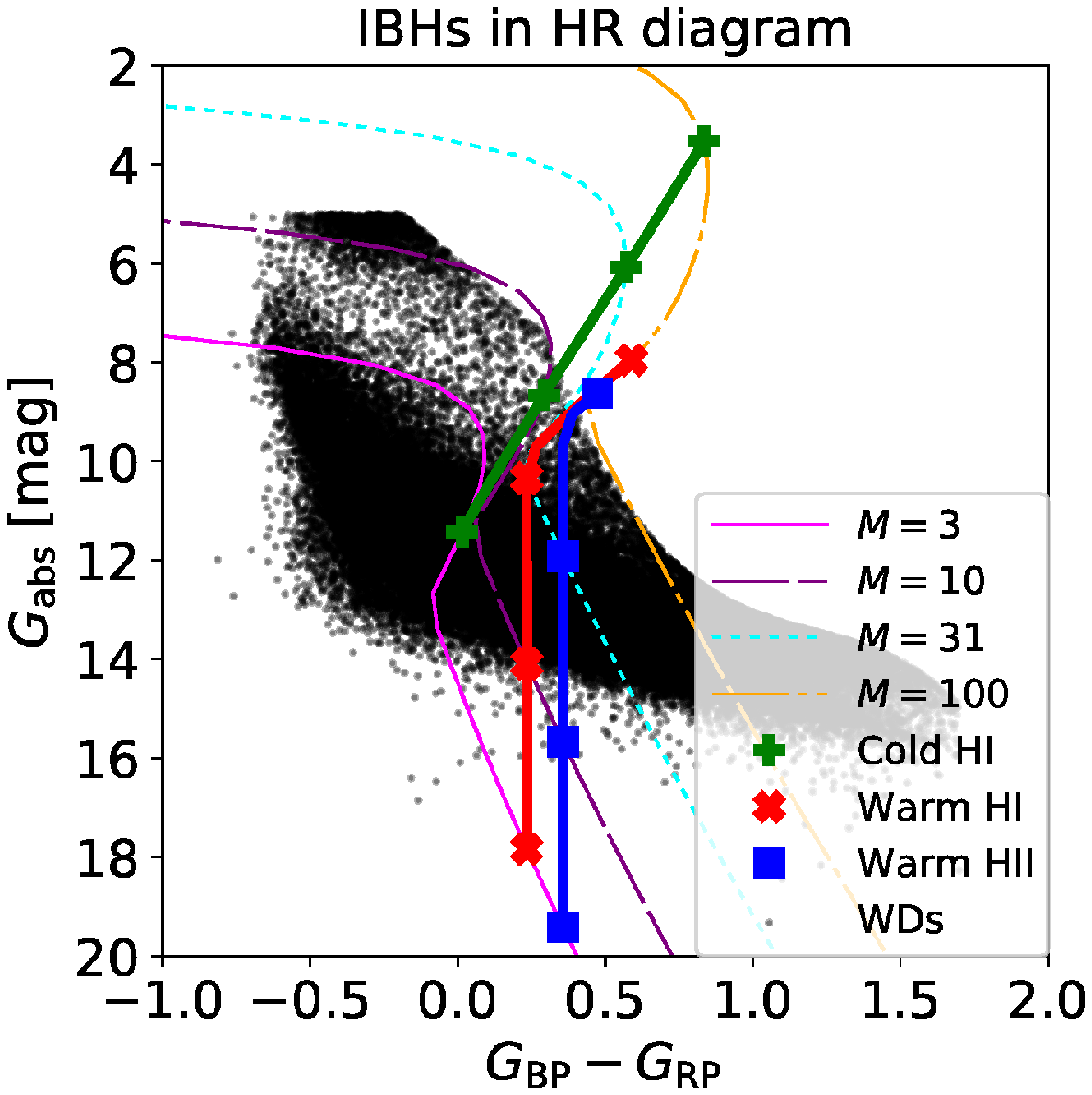}
    \includegraphics[width=\linewidth]{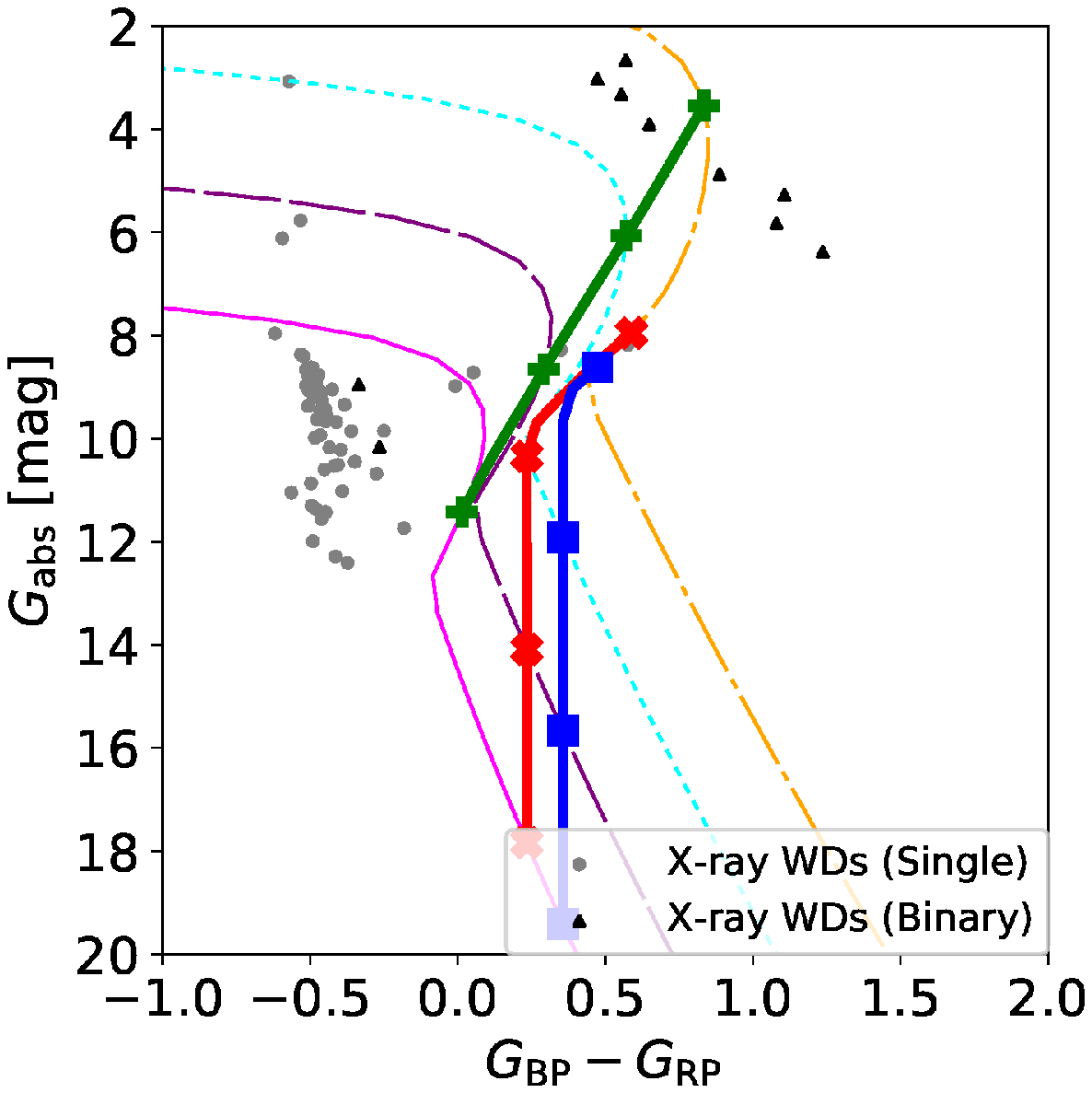}
    \caption{Regions where IBH-MADs occupy in the HR diagram. The thick lines with symbols indicate the IBH-MADs in cold HI (green plus), warm HI (red cross), and warm HII (blue square). The thick lines are obtained by changing the IBH mass. The thin lines depict the sequences of an IBH with various values of the mass accretion rate. In the top panel, we also plot WD candidates detected by Gaia~\citep{2019MNRAS.482.4570G}.
    In the bottom panel, we plot X-ray emitting WDs (grey circles and black triangles) in \citet{1996A&A...316..147F}, and their classification is given by \citet{2017ASPC..509....3D}.   }
    \label{fig:HR}
   \end{center}
  \end{figure}

  \begin{figure}
   \begin{center}
    \includegraphics[width=\linewidth]{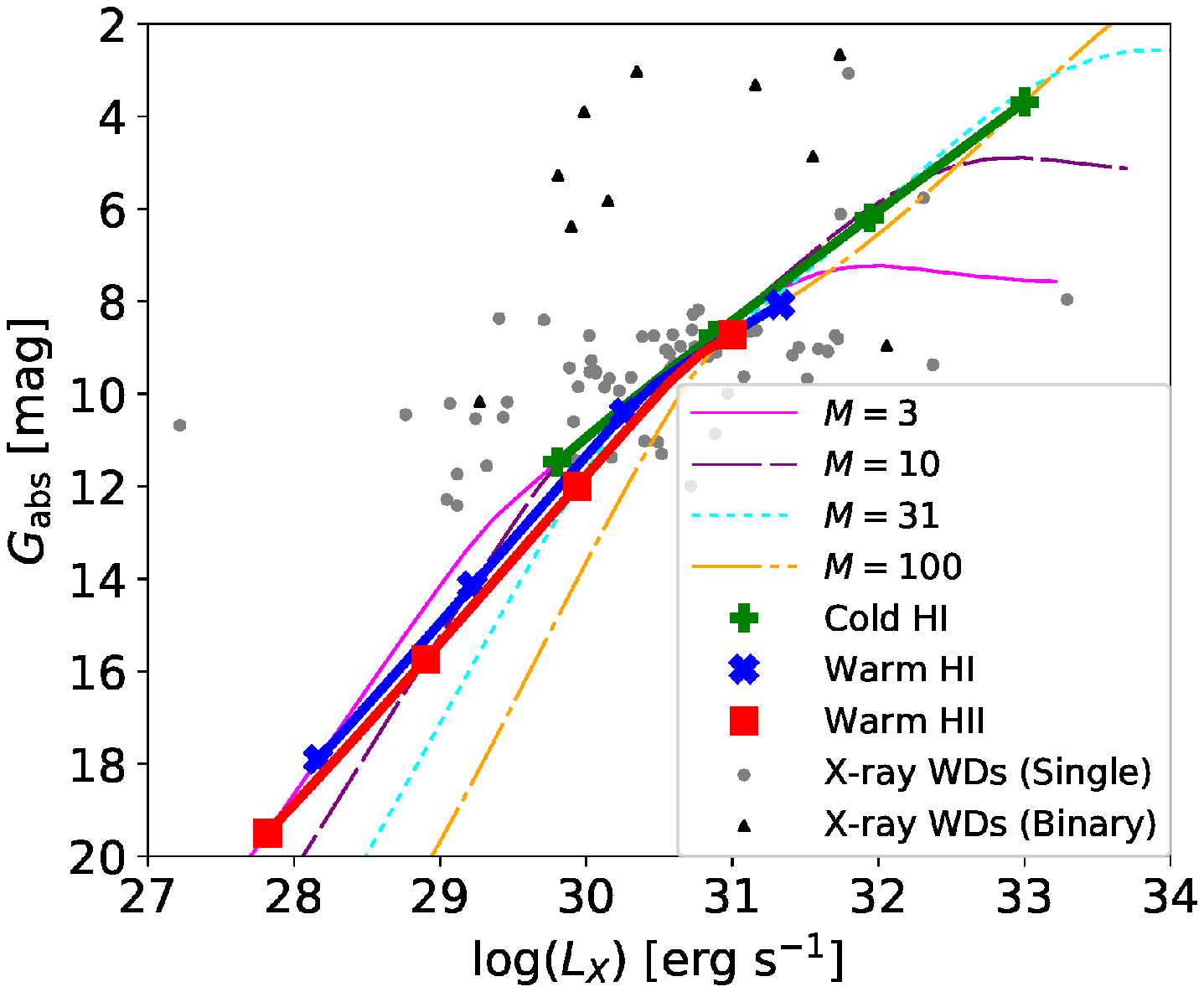}
    \caption{The relation between X-ray luminosity and Gaia G-band absolute magnitude for IBHs (thin and thick lines) and X-ray emitting WDs (grey circles and black triangles).}
    \label{fig:LxLgaia}
   \end{center}
  \end{figure}

 First, we roughly estimate the number of IBHs that can be detected by Gaia or eROSITA. We estimate the detection horizon, $d_{i,\rm det}={\rm min}(\sqrt{L_{i,\rm band}/(4\pi f_{i,\rm sen})},~d_{\rm max})$, where $L_{i,\rm band}$ is the luminosity in the energy band for the detector (330 nm -- 1050 nm for Gaia; 0.2 keV -- 2.3 keV for eROSITA), $f_{i,\rm sen}$ is the sensitivity of the detector (20 mag for Gaia DR5 and $1.1\times10^{-14}\rm~erg~s^{-1}~cm^{-2}$ for the eROSITA four-year survey), and $d_{\rm max}$ is the maximum distance. We set $d_{\rm max}=2$ kpc because Gaia cannot precisely measure the parallax for faint sources and the extinction and attenuation may affect the detectability.

The expected number of detectable IBH candidates can be estimated to be 
\begin{equation}
\mathcal{N}_{\rm det}(M)\sim M\frac{dN_{\rm IBH}}{dMdV}\xi_0{\rm~min}\left(\frac{4\pi}{3}d_{i,\rm det}^3,~2\pi H_{\rm ISM}d_{i,\rm det}^2\right),
\end{equation}
where $H_{\rm ISM}$ is the scale height of each ISM phase (see Table \ref{tab:ism}) and $dN_{\rm IBH}/(dMdV)$ is the number of IBHs per unit mass and volume. We assume a simple power-law mass spectrum with spectral index suggested by the gravitational-wave data: $dN_{\rm IBH}/dM\propto M^{-\gamma}$ with $\gamma\sim2.6$ \citep{2021ApJ...913L...7A}. We consider the mass range of IBHs of $3.2~M_\odot\le M\le50~M_\odot$. The mass-integrated number density of IBHs is set to be $dN_{\rm IBH}/dV=\int(dN/dMdV)dM=10^5\rm~kpc^{-3}$, which is roughly consistent with N-body simulations by \citet{2018MNRAS.477..791T}. The resulting values of $\mathcal{N}_{\rm det}$ are plotted in the left panel of Figure \ref{fig:dndm}. We can see that both eROSITA and Gaia will detect $\sim10^3$ IBHs in cold HI medium in a broad mass range. Several hundreds (around a hundred) of low-mass ($M\sim5~M_\odot$) IBHs in warm HI (warm HII) medium can be discovered by eROSITA, while Gaia can detect only $\sim10$  ($\sim1$) low-mass IBHs in warm HI (warm HII) medium. More than a thousand high mass IBHs in warm HI can be detected by both Gaia and eROSITA. We should note that both the mass spectrum and volumetric density of IBHs are very uncertain. The data by OGLE microlensing surveys suggest a flatter mass spectrum of IBHs with $\gamma\simeq0.92$ \citep{2021arXiv210713697M}.
Also, the Sun is located in a local bubble \citep{2011ARA&A..49..237F}, which may decrease the detectable number of IBHs within $\sim100$ pc. 

The sensitivity of the ROSAT All-Sky Survey (RASS) is $\sim10^{-13}\rm~erg~s^{-1}~cm^{-2}$ \citep{2016A&A...588A.103B}, which is an order of magnitude lower than that of eROSITA. RASS should detect 0.01 times less IBH candidates than eROSITA, which should contain $\sim10$ low-mass IBH candidates. This number is similar to that of RASS unidentified sources in the northern sky \citep{1999A&A...350..743K}, and thus, our model is consistent with the currently available X-ray data. 

Next, we discuss strategy to identify IBH candidates. The Hertzsprung-Russell (HR) diagram is useful to classify the objects. Figure \ref{fig:HR} exhibits the regions where IBH-MADs occupy in the HR diagram with our reference parameters. We can see that low-mass IBH-MADs in the warm media are located at a fainter and bluer region than the white dwarf (WD) cooling sequence. Ultra-cool WDs and neutron stars (NS), including both pulsars and thermally emitting NSs, can be located in the same region. We can utilize the X-ray feature to distinguish IBH-MADs from them. Pulsars and thermally emitting NSs have high values of X-ray to optical luminosity ratio, $L_X/L_{\rm opt}\gg1$ \citep{2014RPPh...77f6901B,2011ApJ...736..117K}, while low-mass IBHs exhibit $L_X/L_{\rm opt}\sim1$ as discussed in Section \ref{sec:spectrum}. In addition, the expected number of detectable isolated NSs are lower than that of IBH-MADs \citep{2021arXiv210604846T}. Isolated WDs may emit X-rays, but the X-ray emitting WDs detected by RASS \citep{1996A&A...316..147F,2009ApJS..181..444A} are bluer and more luminous in G bands than low-mass IBHs, as shown in the bottom panel of Figure \ref{fig:HR} and Figure \ref{fig:LxLgaia}. Ultra-cool WDs are unlikely to emit bright X-rays. Since eROSITA can detect almost all IBHs detected by Gaia, we will be able to identify good low-mass IBH candidates using Gaia and eROSITA data.

High-mass IBHs of $M\gtrsim50M_\odot$ in warm HI/warm HII or medium-mass ($M\sim10~M_\odot$) IBHs in cold HI are located in a redder and brighter region of the WD cooling sequence. This region might be contaminated by binaries consisting of a WD and a main-sequence star, as seen in the bottom panel of Figure \ref{fig:HR}. They can emit X-rays through the magnetic activity, and the values of $L_X/L_{\rm opt}$ are also similar to the IBH-MADs (see Figure \ref{fig:LxLgaia}). Nevertheless, we can discriminate them by multi-band photometric observations. WD-star binaries are expected to have a two-temperature blackbody spectrum in optical bands, while IBH-MADs should exhibit a single smooth component of thermal synchrotron spectrum.

IBH-MADs are likely variable within dynamical timescale, and thus, we should expect strong intra-night variability compared to WDs. 
ULTRACAM can detect sub-second variability, which is a smoking-gun signal to distinguish IBH-MADs from WDs.
Also, IBH-MADs should not show any absorption and emission lines. Therefore, spectroscopic and photometric follow-up observations in both X-ray and optical bands may be useful to distinguish IBHs from WDs. Considering the limiting magnitude of Gaia ($\sim$20 mag), IBH candidates are detectable with 2-meter telescopes for photometric follow-up observations, or with 4-meter telescopes for spectroscopic observations. Detailed soft X-ray spectra of IBH-MADs can be obtained by current and near future X-ray satellites, including Chandra (see Figure \ref{fig:spe}) and XRISM \citep{2020arXiv200304962X}. Besides, IBH-MADs can be detected by future hard X-ray (FORCE: \citealt{2018SPIE10699E..2DN}) and MeV gamma-ray (e.g., GRAMS: \citealt{2020APh...114..107A}) detectors as seen in Figure \ref{fig:spe}, which will strongly support our IBH-MAD scenario. If IBH-MADs are bright enough as in the left panel of Figure \ref{fig:spe}, NuSTAR is also able to detect them.

\section{Discussion} \label{sec:discussion}

We have discussed strategy to identify IBHs based on multi-wavelength emission model of MADs.  Thermal and non-thermal electrons in MADs emit optical and X-ray signal, respectively, which are detectable by Gaia and eROSITA. We can discriminate IBH-MADs from other objects using $L_X/L_{\rm opt}$ and the HR diagram. Hard X-ray and MeV gamma-ray detections will enable us to firmly identify IBHs.

The mass accretion rate onto IBH-MADs can be lower than our reference parameter set due to a lower $\lambda_w$\footnote{We use the most optimistic value, $\lambda_w=1$.} (outflows/convection) or a higher $v_k$. In order to check the detection prospects with lower values of $\dot{M}_\bullet$, we also calculate emission from IBH-MADs with various values of $\dot{M}_\bullet$. The thin lines in Figure \ref{fig:HR} show the tracks for IBH-MADs of a fixed mass with various $\dot{M}_\bullet$.
On each line, IBH-MADs with higher values of $\dot{M}_\bullet$ reside an upper region, and we can see three branches in each line. For the low-$\dot{M}_\bullet$ branch, the electron temperature is independent of $\dot{M}_\bullet$ owing to inefficient radiative cooling. The magnetic fields are stronger and the synchrotron frequency is higher for IBH-MADs with a higher $\dot{M}_\bullet$, and thus, IBH-MADs are bluer when more luminous. In the medium-$\dot{M}_\bullet$ branch, the radiative cooling is efficient enough to balance the heating. In this case, the electron temperature is lower for a higher $\dot{M}_\bullet$, making the branch redder when more luminous. For the high $\dot{M}_\bullet$ branch, the Gaia band is optically thick for the SSA process. Then, the spectral shape below the absorption frequency is given by the Rayleigh-Jeans spectrum, exhibiting a very bluer color when more luminous.
The middle and right panels of Figure \ref{fig:dndm} indicate the expected detection number as a function of $\lambda_w$. Gaia cannot detect  10-$M_\odot$ IBHs in warm media for $\lambda_w<0.1$, while 10-$M_\odot$ IBHs in cold HI can be detectable by Gaia for $\lambda_w\gtrsim0.01$. eROSITA can still detect several tens of 30-$M_\odot$ IBH-MADs even for $\lambda_w\sim0.001$. Optical followup observations of eROSITA unidentified sources will be important to identify IBHs for the cases with $\lambda_w<0.01$.
Observations of nearby low-luminosity AGNs, including Sgr A*, indicate $\lambda_w\sim0.001-0.01$ \citep[e.g.,][]{2018MNRAS.476.1412I}, while the values for IBHs are currently unclear. 

The value of electron heating fraction, $f_e$, is also uncertain. The electron heating prescription by \cite{2018ApJ...868L..18H} suggests $f_e\sim0.1$. This leads IBH-MADs to redder regions in the HR diagram ($G_{\rm BP}-G_{\rm RP}\sim1-2$) . Also, IBH-MADs is luminous in X-rays compared to the optical band as $L_X/L_{\rm opt}\propto f_e^{-1}$ (see Equations (\ref{eq:Lsyn}) and (\ref{eq:Lx})). Despite these uncertainties, our strategy of IBH identification should still work even with low values of $f_e$, because faint and red WDs are very unlikely to emit X-rays. Therefore, we suggest to search for IBH candidates in a broad region of the HR diagram using X-ray data.

The spectral index of reconnection acceleration may be softer than our assumption of $s_{\rm inj}=1.3$, but our conclusions are unaffected as long as we use $s_{\rm inj}\lesssim2$. Recent 3D particle-in-cell (PIC) simulations support a hard spectral index of $s_{\rm inj}\sim1$ \citep{2021arXiv210500009Z}, and long-term 2D PIC simulations suggest $s_{\rm inj}\simeq2$ \citep{2018MNRAS.481.5687P}. These results support our assumption. In contrast, other 2D PIC simulations indicate much softer spectra ($s_{\rm inj}\sim3-4$) at a high-energy range of $E_e \gtrsim \sigma_B m_ec^2$ \citep{2018ApJ...862...80B,2018MNRAS.473.4840W,2021ApJ...912...48H}. If particle acceleration by magnetic reconnections results in a soft spectral index, the subsequent stochastic acceleration by turbulence is necessary \citep{2018PhRvL.121y5101C} to emit strong X-ray signals.

The IBHs may emit radio and sub-mm signals. Although IBH-MADs cannot produce detectable radio signals, compact jets are highly likely launched by MADs \citep{TNM11a}. Such jets may produce radio signals as detected from a few quiescent X-ray binaries \citep[e.g.,][]{2009MNRAS.399.2239H,2019MNRAS.488..191G}. The radio luminosity correlates with X-ray luminosity in the low-hard state: $L_R\sim1.4\times10^{29}L_{X,35}^{0.61}\rm~erg~s^{-1}$ \citep{2014MNRAS.445..290G}. If we extrapolate this relation to the regime of IBHs, the radio flux can be estimated to be $F_R\sim0.63~L_{X,28}^{0.61}(d/100\rm~pc)^{-2}$ mJy. Thus, the signals from compact jets are detectable by current radio telescopes, such as ALMA and VLA. These signals may be detectable by ongoing radio surveys, such as Very Large Array Sky Survey \citep[VLASS;][]{2020PASP..132c5001L}, ThunderKAT \citep{ThunderKAT17a}, and ASKAP Survey for Variable and Slow Transients \citep[VAST;][]{2013PASA...30....6M}. Current observational data of quiescent X-ray binaries are not precise enough to obtain the $L_R-L_X$ relation. $L_R$ might decrease more rapidly than $L_X$ in a highly sub-Eddington regime \citep{2020ApJ...889...58R}, which leads to 1--2 orders of magnitude lower radio flux. Even in this case, the next-generation radio facilities will be able to detect radio signals from IBHs, which will provide another support of the IBH-MAD scenario.

\begin{acknowledgments}
We thank the anonymous reviewer for useful comments. This work is partly supported by JSPS Research Fellowship and KAKENHI No. 19J00198 (S.S.K.), and 20K04010, 20H01904 (K.K).
\end{acknowledgments}

\bibliography{ssk}
\bibliographystyle{aasjournal}



\end{document}